\documentclass[twocolumn]{revtex4}
\usepackage{graphicx}% Include figure files
\usepackage{amsmath,amsbsy,amssymb}
\usepackage{bm}% bold math
\usepackage{mathrsfs,color}

\newcommand{\imag}{\mathrm{Im}\,}
\newcommand{\real}{\mathrm{Re}\,}

\newcommand{\imu}{\mathrm{i}}
\newcommand{\epn}{\mathrm{e}}

\newcommand{\ua}{\uparrow}
\newcommand{\da}{\downarrow}
\newcommand{\dg}{\dagger}
\newcommand{\la}{\langle}
\newcommand{\ra}{\rangle}
\newcommand{\al}{\alpha}
\newcommand{\sg}{\sigma}
\newcommand{\gm}{\gamma}
\newcommand{\ep}{\varepsilon}

\begin{document}

\title{
Tunneling and Josephson effects in odd-frequency superconductor junctions: \\
A study on multi-channel Kondo chain
}

\author{Shintaro Hoshino$^{1}$, Keiji Yada$^2$ and Yukio Tanaka$^2$}

\affiliation{
$^1$Department of Basic Science, The University of Tokyo, Meguro, Tokyo 153-8902, Japan
\\
$^2$Department of Applied Physics, Nagoya University, Nagoya, 464-8603, Japan
}

\date{\today}

\begin{abstract}
Junction systems of odd-frequency (OF) superconductors are investigated based on a mean-field Hamiltonian formalism.
One-dimensional two-channel Kondo lattice (TCKL) is taken as a concrete example of OF superconductors.
Properties of normal and Andreev reflections are examined in a normal metal/superconductor junction.
Unlike conventional superconductors, 
normal reflection is always present
due to the normal self energy that necessarily appears in the present OF pairing state.
The conductance reflects the difference between repulsive and attractive potentials located at the interface, which is in contrast with the 
preexisting superconducting junctions.
Josephson junction is also constructed by connecting TCKL with the other types of superconductors.
The results can be understood from symmetry of the induced Cooper pairs at the edge in the presence of spin/orbital symmetry breaking.
It has also been demonstrated that the symmetry argument for Cooper pairs is useful in explaining Meissner response in bulk.
\end{abstract}

\maketitle

\section{Introduction}

Odd-frequency (OF) superconductivity \cite{berezinskii74, kirkpatrick91, belitz92,balatsky92,emery92}, which is characterized by pair potential or pair amplitude with odd functions in time and frequency, has been recognized as a candidate of intriguing quantum states of matter.
While its realization has been theoretically proposed in a variety of systems \cite{Vojta,Fuseya,Shigeta,Hotta,Otsuki} there has been a long-standing problem: a thermodynamic instability arises if we apply a conventional approach to OF superconductivity which has succeeded in describing ordinary even-frequency (EF) superconductors \cite{coleman93, heid95}.
At the same time, the sign of electromagnetic response function is reversed from the usual diamagnetic one, indicating a paramagnetic Meissner response and negative superfluid density.
Therefore, its realization in bulk of condensed matter has been questioned.
On the other hand, without such difficulties the OF superconductivity can exist as a surface state.
While the ordinary EF superconductivity is dominant in bulk, the OF pairing state is present as an induced state \cite{bergeret05, tanaka07, higashitani12, tanaka12}.
It has been reported that the paramagnetic Meissner response \cite{higashitani97,tanaka05,yokoyama11,Asano15} 
is observed in this induced OF pairing state \cite{Robinson}.

Recent theoretical studies show that the OF pairing can also be stabilized in a bulk, if we reconsider some conditions which are usually assumed in the theory of conventional superconductors.
Namely, the sign arising from the OF pair potential $\Delta(-\omega) = - \Delta(\omega)$, which causes the thermodynamic instability, can be canceled by introducing an additional sign.
One of the solutions is to re-examine the conjugate relations of the frequency-dependent pair potential \cite{belitz99, solenov09, kusunose11}.
Using a path-integral formalism, it has been shown  in Refs.~\onlinecite{solenov09} and \onlinecite{kusunose11} that the problem can be resolved by using an unusual conjugate relation for pair potential, and consequently a description based on Hermitian mean-field Hamiltonian is impossible.
Subdominant even-frequency pairings induced in inhomogeneous systems such as surface or defect have been classified for this kind of OF pairing based on symmetry arguments \cite{asano14}.
With this situation, however, recently Josephson junctions have also been studied by Fominov {\it et al.} and peculiar properties are revealed
\cite{fominov15}.
Whereas a real current is obtained for the junction with EF superconductor, the current becomes imaginary if we make a junction with paramagnetic OF superconductivity realized at e.g. the edges of $p$- or $d$-wave superconductors.
Furthermore, an electromagnetic response function shows that the superfluid weight becomes complex number when these diamagnetic and paramagnetic odd-frequency pairings coexist \cite{fominov15}.
These unphysical behaviors at least indicate that the conventional approach fails in describing the coexistence of the above OF superconductor.

On the other hand, it has also been established that there is another type of stable OF superconductivity, which can be described in the mean-field theory with Hermitian Hamiltonian.
Here, the additional minus sign to resolve the thermodynamic problem comes from a spatially oscillating phase of the pair amplitude \cite{coleman93}, which is called staggered pairing \cite{martisovits98,heid95-2}.
The existence of  staggered OF pairing has been clearly demonstrated in the two-channel Kondo lattice (TCKL) \cite{hoshino14-1,hoshino14-2}.
In this paper, for a deeper understanding of this type of OF superconductors, we take TCKL as a concrete example and explore novel properties of the junction systems to clarify the difference from already known superconductors.

\begin{table}[t]
\begin{tabular}{c|c||c|c|c|c}
\hline
modulation & identifier & time & spin & space & orbital
\\
\hline
uniform & ESEE &$+$&$-$&$+$&$+$
\\
 & ESOO &$+$&$-$&$-$&$-$
\\
 & ETOE  &$+$&$+$&$-$&$+$
\\
 & ETEO  &$+$&$+$&$+$&$-$
\\
 & OTEE  &$-$&$+$&$+$&$+$
\\
 & OTOO &$-$&$+$&$-$&$-$
\\
 & OSOE &$-$&$-$&$-$&$+$
\\
 & OSEO &$-$&$-$&$+$&$-$
\\ \hline
staggered & ESEE(+OSOE)  &$+(-)$&$-$&$+(-)$&$+$
\\
 & ESOO(+OSEO)  &$+(-)$&$-$&$-(+)$&$-$
\\
 & ETOE(+OTEE) &$+(-)$&$+$&$-(+)$&$+$
\\
 & ETEO(+OTOO)  &$+(-)$&$+$&$+(-)$&$-$
\\
 & OSOE(+ESEE)  &$-(+)$&$-$&$-(+)$&$+$
\\
 & OSEO(+ESOO)  &$-(+)$&$-$&$+(-)$&$-$
\\
 & OTEE(+ETOE) &$-(+)$&$+$&$+(-)$&$+$
\\
 & OTOO(+ETEO)  &$-(+)$&$+$&$-(+)$&$-$
\\
\hline 
\end{tabular}
\caption{
Classification of uniform and staggered pairs.
The sign represents even or odd character of the exchange symmetry.
The symmetry in bracket is a secondarily induced pair from a purely OF or EF pair potential.
}
\label{tab:list}
\end{table}

Superconducting states are classified by internal structures of the Cooper pairs \cite{black-schaffer13, asano15, triola15}:
even (E) or odd (O) in time (frequency),
triplet (T) or singlet (S) in spin space,
even (E) or odd (O) in real space (momentum), and
even (E) or odd (O) in orbital space. 
To be consistent with Fermi-Dirac statistics, 
the exchange symmetry of these indices must be odd in total.
Hence, we can have eight classes of symmetry of Cooper pair 
labeled by ESEE, ESOO, ETOE, ETEO, OTEE, OTOO, OSOE, and OSEO.
We additionally consider uniform and staggered spatial modulation patterns for pairing states.
There is a similar state called pair density wave (PDW) which also has a spatially oscillating pair amplitude \cite{podolsky03,chen04,ohkawa06,seo08,agterberg08, berg10,yoshida12,cho14}.
If the modulation appears in a staggered manner, PDW is identified as the staggered pairing or $\eta$-pairing \cite{yang89,singh91,boer95,japaridze01,zhai05}.
These pairing states are characterized by a finite center-of-mass momentum.
Recently, the Amperean pairing, where the electrons with same momentum are paired, is theoretically proposed in the context of pseudo gap phase in cuprate 
\cite{lee14, baireuther15}.
This state also carries a finite center-of-mass momentum.
While these concepts have been developed without relation to OF pairing, here we demonstrate that these spatially modulated states are closely 
connected to it.

With spatial modulations of pair amplitudes, OF and EF components mix.
Table~\ref{tab:list} summarizes the pairing states.
Even when OF and EF pair amplitudes coexist, we can clearly define the OF superconductivity in some situations, since the pair {\it potential} 
can have only OF component.
Indeed this is the case realized in TCKL as will be discussed in this paper.
It will also be demonstrated that the symmetry argument is quite useful to discuss Meissner response of OF superconductors and Josephson junctions.

In this paper we discuss the staggered OF superconductivity in TCKL based on a mean-field Hamiltonian.
The model and bulk properties are discussed in Secs.~II and III, which can be applied to systems in arbitary dimensions.
Junctions in one dimension are made using OF superconductors and normal metal to investigate characteristics of Andreev reflection (Sec.~IV) and Josephson current (Sec.V).
We summarize the obtained results in Sec.~VI.

\section{Model}

We begin with the two-channel Kondo lattice \cite{jarrell96, cox98} whose Hamiltonian is given by
\begin{align}
\mathscr{H}_{\rm TCKL} &= \sum_{ij\sg m } (-t_{ij} - \mu \delta_{ij}) c^\dg_{i\sg m } c_{j\sg m }
\nonumber \\
&\ \ \ + \frac{J}{2} \sum_{i\sg m  m '} \bm \tau_i \cdot c^\dg_{i\sg m } \bm \sg_{ m  m '} c_{i\sg m '}
\end{align}
where $c_{i\sg m }$ ($c^\dg_{i\sg m }$) is an annihilation (creation) operator of electrons at site $i$ with spin $\sg=\ua,\da$ and orbital $ m =1,2$.
We define the $2\times 2$ Pauli matrix $\bm \sg$ and the localized pseudospin operator $\bm \tau_i$ at site $i$.
This model is known as an effective model for $f$-electron systems with non-Kramers configuration, and can be applicable to some Pr- and U-based compounds \cite{cox98}.
Since in $f$-electron systems the spin and orbital are coupled and the total angular momentum is a good quantum number, the index $\sg$ physically describes a Kramers index, which is connected by the time-reversal symmetry, and the orbital $ m $ a non-Kramers index.
For simplicity we call $\sg$ ``spin'', and $m$ ``orbital'' in the following.
The hopping integrals are given by $t_{ij}=t$ for the nearest neighbor sites and otherwise zero.
In Secs.~II and III we consider a general simple bipartite lattice such as a cubic lattice and do not restrict ourselves to a one-dimensional chain.

In the previous study \cite{hoshino14-1}, TCKL has been investigated by the dynamical mean-field theory, and the existence of the OF superconductivity has been numerically demonstrated.
The corresponding effective mean-field theory has also been successfully constructed \cite{hoshino14-2}, which qualitatively describes the pairing state at low temperatures.
In this framework, we first rewrite the localized pseudospin-operator $\bm \tau_i$ by introducing pseudofermion as
$\bm \tau_i = \frac 1 2 \sum_{ m  m '} f^\dg_{i m } \bm \sg_{ m  m '} f_{i m '}$
with the local operator constraint $\sum_ m  f^\dg_{i m } f_{i m } = 1$ at every site.
In the mean-field theory, this constraint is satisfied only in the mean value: $\sum_ m  \la f^\dg_{i m } f_{i m } \ra = 1$.
The interaction term is decoupled by the mean-field approximation. 
This procedure is dependent on the spin:
\begin{align}
J \bm \tau_i \cdot c^\dg_{i\ua m }\bm \sg_{ m  m '} c_{i\ua m '}
&\longrightarrow V \delta_{ m  m '} f^\dg_{i m } c_{i\ua m' } + {\rm H.c.}
\\
J \bm \tau_i \cdot c^\dg_{i\da m }\bm \sg_{m m '} c_{i\da m '}
&\longrightarrow V \epn^{\imu \bm Q \cdot \bm R_i}\epsilon_{ m  m '} f^\dg_{i m } c^\dg_{i\da m '} + {\rm H.c.}
\end{align}
where we have defined the antisymmetric tensor $\epsilon = \imu \sg^y$ and the staggered ordering vector $\bm Q= (\pi,\pi, \cdots)$ with the lattice constant being unity.
The vector $\bm R_i$ represents the position of site $i$.
The mean-field potential $V$ is determined by the self-consistent equation
\begin{align}
V &= - \frac{3J}{4} \la c^\dg_{i\ua m } f_{i m } \ra
= \frac{3J}{4} \la c_{i\da m } f_{i m' } \ra \epsilon_{mm'} \epn^{\imu \bm Q\cdot \bm R_i}
\label{eq:def_v}
\end{align}
Here the middle part (diagonal quantity) 
and last part (off-diagonal quantity) can be different in general. 
However, the difference causes the channel symmetry breaking \cite{hoshino14-2} and is energetically unfavorable away from half filling \cite{hoshino14-1}.
Hence we take the same value as in Eq.~\eqref{eq:def_v}, which corresponds to 
the U(1) symmetry breaking without any other spontaneous symmetry breaking.
This point can also be justified because the dynamical mean-field theory, which takes full account of local correlation effects, shows the qualitatively same results at low energies and low temperatures.
For simplicity, in the following of this paper we neglect the self-consistency of the equation, and we take the mean-field $V$ as an input parameter.

At half filling, this pair is likely to be formed since both the wave vectors $\bm k$ and $-\bm k-\bm Q$ can be placed on the Fermi surface.
In fact, this pairing state can exist away from half filling, even though the direct pair between $\bm k$ and $-\bm k-\bm Q$ cannot be formed on the Fermi surface of conduction electrons.
This is because the condensation energy is gained not by the pair between conduction electrons, but by the pair between conduction electron and pseudofermion as shown in Eq.~\eqref{eq:tckl_mf}.
The energy gain is an increasing function of a size of Fermi surface, and hence the transition temperature becomes maximum at half filling \cite{hoshino14-1}.

From the expression of Eq.~\eqref{eq:def_v} one might think that the present pairing is an even-frequency pairing, since there is no time dependence in the pairing amplitude.
However, the pseudofermions are nothing but {\it virtual} degrees of freedom, and must be traced out to evaluate physical quantities.
This is because the fermionic operator $f_{i m}$ is introduced to describe the localized pseudospin and such fermion does not exist in the original Hamiltonian.
Accordingly experimentally measurable physical quantities should not directly include these pseudofermions.
The relevant order parameter is then given by a time-dependent pairing amplitude composed only of conduction electrons as seen below.

The mean-field Hamiltonian is explicitly written down as \cite{hoshino14-2}
\begin{align}
&\mathscr{H}_{\rm TCKL}^{\rm MF} 
= \sum_{ij\sg m } (-t_{ij} - \mu \delta_{ij}) c^\dg_{i\sg m } c_{j\sg m }
\nonumber \\
&
+ V\sum_{i m  m '} (\delta_{ m  m '} f^\dg_{i m } c_{i\ua m' }
+ \epsilon_{ m  m '} \epn^{\imu \bm Q\cdot \bm R_i} f^\dg_{i m }c^\dg_{i\da m' } + {\rm H.c.})
\label{eq:tckl_mf}
\end{align}
Here the phase of the pair amplitude is fixed.
We can also consider self energies from Green functions \cite{hoshino14-2}.
After tracing out the pseudofermion degrees of freedom, the normal and anomalous self energies (pair potential) for conduction electrons are given by
\begin{align}
\Sigma_i(z) &= \frac{V^2}{z}
\label{eq:normal_self}
\\
\Delta_i(z) &= \frac{V^2 \epn^{\imu \bm Q \cdot \bm R_i}}{z} = \frac{V^2 \cos (\bm Q \cdot \bm R_i)}{z}
\label{eq:anomalous_self}
\end{align}
respectively, where $z=\ep+\imu \eta$ for real frequencies ($\eta$ is positive infinitesimal) and $z=\imu \ep_n=(2n+1)\pi \imu T$ for Matsubara frequencies.
This expression explicitly demonstrates the realization of staggered OF superconductivity: the pair potential has only odd-frequency component.
As seen in Eq.~\eqref{eq:anomalous_self} the staggered pairing is regarded as both ``Fulde-Ferrell (FF)'' and ``Larkin-Ovchinnikov (LO)'' states \cite{fulde64,larkin65}.
While the spatial modulation is slowly varying in FFLO pairing, 
the spatial oscillation in TCKL is much faster than that in FFLO state.

We show that the staggered property of pair amplitude can be removed by the local gauge transformation only for $\sg=\da$ defined by
$c_{i\da m } \longrightarrow c_{i\da m } \epn^{\imu \bm Q\cdot \bm R_i}$.
With this the Hamiltonian is transformed as
\begin{align}
&\mathscr{H}_{\rm TCKL}^{\rm MF}
\longrightarrow \tilde{\mathscr{H}}_{\rm TCKL}^{\rm MF}
= \sum_{ij\sg m } (-t_{ij} \sg^z_{\sg\sg} - \mu \delta_{ij}) c^\dg_{i\sg m } c_{j\sg m }
\nonumber \\
& + V\sum_{i m  m '} (\delta_{ m  m '} f^\dg_{i m } c_{i\ua m' }
+ \epsilon_{ m  m '} f^\dg_{i m }c^\dg_{i\da m' } + {\rm H.c.})
. \label{eq:transf_ham}
\end{align}
Thus the staggered nature is completely washed away for infinite system.
At the same time the sign of the hopping of 
an electron with $\da$-spin is reversed.
While we have emphasized that the staggered nature is important for the thermodynamically stable OF superconductivity, the present argument shows that it is not the only way.
Namely the effect from the staggered phase can be replaced by the sign difference between hoppings of electrons with the two spins.
With this Hamiltonian, the normal and anomalous self-energies are given respectively by
\begin{align}
\Sigma_{i}(z) &= \Delta_i(z) = \frac{V^2}{z}
\label{eq:normal_self_mod}
\end{align}
Note that we now obtain the uniform OF state, with spin-symmetry breaking.

\section{Cooper Pairs Formed in TCKL} \label{sec:pair}

Before we study junction systems, let us discuss the symmetry 
of Cooper pairs formed in bulk of TCKL.
The time-dependent pairing amplitude defined by
$ F_{\bm k\bm k'\sg\sg'mm'}(\tau) = - \la T_\tau c_{\bm k\sg m} (\tau) c_{\bm k'\sg'm'} \ra$ 
has the following structure \cite{hoshino14-2}:
\begin{align}
F_{\bm k\bm k'\sg\sg'mm'}(\tau)= \epsilon_{\sg\sg'} \epsilon_{mm'} \delta_{-\bm k-\bm Q, \bm k'}
{\cal F}_{\bm k} (\tau)
, \label{eq:pair_form}
\end{align}
where $T_\tau$ is an imaginary-time ordering operator and $O(\tau) = \epn^{\tau \mathscr{H}} O \epn^{-\tau \mathscr{H}} $.
Eq.~\eqref{eq:pair_form} means the staggered spin-singlet orbital-singlet pair.
From the Fermi-Dirac statistics we have the relation 
$F_{\bm k\bm k'\sg\sg'mm'}(\tau)=-F_{\bm k'\bm k\sg'\sg m' m}(-\tau)$, 
which leads to
\begin{align}
{\cal F}_{-\bm k-\bm Q}(\tau) = - {\cal F}_{\bm k}(-\tau)
. \label{eq:relation}
\end{align}
An explicit form of ${\cal F}_{\bm k} (\tau)$ can be found in Ref.~\onlinecite{hoshino14-2}, but it is not necessary in this paper.

With spatial modulations, EF and OF components should be mixed due to a broken translational invariance. 
To see this explicitly in TCKL, we define 
\begin{align}
F^{\pm}_{\bm k\bm k'\sg\sg'mm'}(\tau)= 
F_{\bm k\bm k'\sg\sg'mm'}(\tau) \pm F_{\bm k\bm k'\sg\sg'mm'}(-\tau)
, \label{eq:def_EF_OF}
\end{align}
where $F^+$ and $F^-$ correspond to the EF and OF pair amplitudes, respectively.
We can show that the exchange of the two wave vectors results in
\begin{align}
F^{\pm}_{\bm k'\bm k\sg\sg'mm'}(\tau)= 
\mp F^{\pm}_{\bm k\bm k'\sg\sg'mm'}(\tau)
\label{eq:pair_symmetry}
\end{align}
As seen from Eq.~\eqref{eq:pair_symmetry}, the exchange symmetry of real space is odd for EF pair and even for OF pair.
For both spin and orbital indices, the exchange symmetries are odd.
Thus the existing pairs in bulk of TCKL are OSEO and ESOO (see also Tab.~\ref{tab:list}).
We note that the primary component is OSEO which arises from purely odd-frequency pair potential in Eq.~\eqref{eq:anomalous_self}, 
and ESOO is a secondarily induced pair.

We comment also on induced pairs in addition to the original pairs OSEO+ESOO when symmetry breaking fields are present.
If we apply the Zeeman field, it mixes up the spin-singlet and spin-triplet.
The induced pairs are then ETEO+OTOO.
In a similar manner, when we apply the external orbital field, which corresponds to a uniaxial pressure, the orbital-odd and even parts are mixed, which causes ESEE and OSOE pairs.
In addition we can also consider another symmetry-breaking field.
This is called spin-orbital field, and breaks both spin and orbital symmetries but their product remain unbroken.
In this case the induced pairs are OTEE+ETOE.
These properties are summarized in Tab.~\ref{tab:induced_pair}, and are important to understanding the Josephson junction as will be discussed in Sec.~V.

\begin{table}[t]
\begin{tabular}{c|c|c}
\hline
 broken symmetry& lifted components & induced pairs
\\
\hline
spin & $(\ua1, \ua2), (\da1, \da2) $ & ETEO+OTOO
\\
orbital & $(\ua1, \da1), (\ua2, \da2) $ & ESEE+OSOE
\\
spin-orbital & $(\ua1, \da2), (\ua2, \da1) $ & OTEE+ETOE
\\
\hline
\end{tabular}
\caption{
Induced pairs in addition to the original OSEO+ESOO pairs when symmetry breaking fields are present in TCKL.
The components in the different brackets are not identical in the presence of the symmetry breaking fields.
}
\label{tab:induced_pair}
\end{table}

For transformed Hamiltonian given in Eq.~\eqref{eq:transf_ham}, the OF pair has a uniform character.
In terms of the classification in Tab.~\ref{tab:list}, the transformed state belongs to uniform OSEO.
Since the spin symmetry is broken in this picture, the uniform pair with ETEO is mixed at the same time.
Thus the secondarily induced pairs are transformed from ESOO to ETEO by the local gauge transformation.

In the following we explain how the OF pairs in TCKL give the ordinary diamagnetic Meissner effect, although the odd-frequency superconductors have long been considered to give a paramagnetic Meissner kernel.
While the numerical calculation \cite{hoshino14-2} shows the diamagnetic response, here we discuss it by focusing on the structure of the Meissner kernel and do not enter the details.
Following the derivation in Ref.~\onlinecite{hoshino14-2}, only the anomalous part contributes to the Meissner kernel $K^{xx}$ which can be written in the form
\begin{align}
K^{xx} &= 
-2e^2 T \sum_{n\bm k\bm k'\sg\sg' m m'} v^x_{\bm k} v^{x}_{\bm k'}
\nonumber \\
&\hspace{10mm} \times
F^\dg_{\bm k\bm k'\sg\sg'mm'}(-\imu\ep_n)
F_{\bm k\bm k'\sg\sg'mm'}(\imu\ep_n)
, \label{eq:meissner_kernel}
\end{align}
where $e$ is an electric charge, and we define the velocity $v^x_{\bm k} = \partial \ep_{\bm k}/\partial k_x$ along the $x$-direction.
We have also introduced the `daggered' anomalous Green function by
$ F^\dg_{\bm k\bm k'\sg\sg'mm'}(\tau) = - \la T_\tau c^\dg_{\bm k\sg m} (\tau) c^\dg_{\bm k'\sg'm'} \ra$.
From Hermiticity of the Hamiltonian, we have the relation
\begin{align}
F^\dg_{\bm k\bm k'\sg\sg'mm'}(-\imu\ep_n) = F^*_{\bm k'\bm k\sg'\sg m'm} (\imu\ep_n)
\end{align}
This relation can be explicitly shown by using the spectral representation.
We assume the inversion symmetry in the original lattice: $ \ep_{-\bm k} =  \ep_{\bm k}$.

Let us consider the conventional 
spin-singlet $s$-wave  (EF) superconductor as a reference.
The anomalous Green's function has the structure in the form
\begin{align}
F^{\rm BCS}_{\bm k\bm k'\sg\sg'}(\imu\ep_n) = \delta_{-\bm k, \bm k'} \epsilon_{\sg\sg'} {\cal F}^{\rm BCS}_{\bm k}(\imu \ep_n).
\end{align}
Here we do not have to know the detailed functional form of ${\cal F}^{\rm BCS}_{\bm k}$.
The orbital degree of freedom is not included here.
The Meissner kernel is then given by
\begin{align}
K^{xx} &= 
-2e^2 T \sum_{n\bm k\sg} v^x_{\bm k} v^{x}_{-\bm k} (\epsilon^2)_{\sg\sg} |{\cal F}^{\rm BCS}_{\bm k}(\imu \ep_n)|^2
\end{align}
For the velocity we have the relation $v^x_{-\bm k} = - v^x_{\bm k}$, which gives the minus sign.
In addition, another sign comes from the spin factor $(\epsilon^2)_{\sg\sg}=-1$, and hence in total the electromagnetic response is diamagnetic: $K^{xx}<0$.
On the other hand, if we had $s$-wave spin-triplet OF superconductivity, there would be no sign from spin-factor.
Hence in this case the sign of the Meissner kernel is reversed to give a paramagnetic response (or sometimes called negative Meissner effect).
For $p$-wave superconductors, the minus sign comes from spatial part, i.e. ${\cal F}^{\rm BCS}_{-\bm k} = - {\cal F}^{\rm BCS}_{\bm k}$, instead of spin part.

Now we consider the kernel in TCKL.
Substituting Eq.~\eqref{eq:pair_form} into Eq.~\eqref{eq:meissner_kernel}, we obtain
\begin{align}
K^{xx} &= 
-2e^2 T \sum_{n\bm k\sg m} v^x_{\bm k} v^{x}_{-\bm k-\bm Q} (\epsilon^2)_{\sg\sg} (\epsilon^2)_{mm} 
\nonumber \\
&\hspace{10mm} \times 
{\cal F}_{-\bm k-\bm Q}^*(\imu \ep_n) {\cal F}_{\bm k}(\imu \ep_n)
\end{align}
Although the factors from spin and orbital parts give the minus sign as $(\epsilon^2)_{\sg\sg}=(\epsilon^2)_{mm}=-1$, the sign operates twice and does not affect the total Meissner kernel.
For the velocity, we have $v^{x}_{-\bm k-\bm Q} = v^x_{\bm k}$ originating from $\ep_{\bm k+\bm Q}+\ep_{\bm k}=0$, which gives no minus sign in contrast to the above $s$-wave spin-singlet superconductor.
We further transform the expression in terms of EF and OF pair amplitudes ${\cal F}^{\pm}_{\bm k}(\imu \ep_n)$ originating from Eqs.~\eqref{eq:pair_form} and \eqref{eq:def_EF_OF}.
Using the relation in Eq.~\eqref{eq:relation}, the final expression is written as
\begin{align}
K^{xx} &= 
2e^2 T \sum_{n\bm k} (v^x_{\bm k})^2 
\left[
|{\cal F}_{\bm k}^+(\imu \ep_n)|^2 - |{\cal F}_{\bm k}^-(\imu \ep_n)|^2
\right]
. \label{eq:tckl_kernel}
\end{align}
Namely the OF pair (OSEO) gives a diamagnetic contribution 
and the EF pair (ESOO) shows a paramagnetic response, 
which is contrary to the standard wisdom.
Although it is not trivial to determine which parts give the dominant contribution, the numerical calculation shows that the OF part is more dominant to give the total diamagnetic response \cite{hoshino14-2}.
This fact implies the importance of the OF pair in TCKL.

The characteristic diamagnetic response by OF pairs in TCKL is closely related to  $v^{x}_{-\bm k-\bm Q} = v^x_{\bm k}$ with finite center-of-mass momentum $\bm Q$.
Otherwise we would have another minus sign from $v^{x}_{-\bm k} = - v^x_{\bm k}$ and then the OF pair gives paramagnetic contribution.
This point has also been numerically demonstrated in Ref.~\onlinecite{hoshino14-2}.

At the end of this section, let us also consider the Meissner kernel in the modified TCKL given by Eq.~\eqref{eq:transf_ham}.
In this case the uniform pair amplitudes have the form
\begin{align}
&F_{\bm k\bm k'\sg\sg'mm'}(\tau) 
\nonumber \\
&=\frac 1 2
 \epsilon_{mm'} \delta_{-\bm k, \bm k'}
\left[
\epsilon_{\sg\sg'} {\cal F}^{-}_{\bm k} (\imu\ep_n) 
+ \sg^x_{\sg\sg'} {\cal F}^{+}_{\bm k} (\imu\ep_n) 
\right]
. \label{eq:pair_sym_mod}
\end{align}
The first and second terms in the right-hand side respectively correspond to OSEO and ETEO pairs.
The Meissner kernel has the form
\begin{align}
K^{xx} &= 
-2e^2 T \sum_{n\bm k\bm k'\sg\sg' m m'} v^x_{\bm k\sg} v^{x}_{\bm k'\sg'}
\nonumber \\
&\hspace{10mm} \times
F^*_{\bm k'\bm k\sg'\sg m'm}(\imu\ep_n)
F_{\bm k\bm k'\sg\sg'mm'}(\imu\ep_n)
, \label{eq:Meissner_mod}
\end{align}
instead of Eq.~\eqref{eq:meissner_kernel}.
The important point here is that the velocity is dependent on spin: $\bm v_{\bm k\ua} = \bm v_{\bm k}$ and $\bm v_{\bm k\da} = - \bm v_{\bm k}$.
Substituting Eq.~\eqref{eq:pair_sym_mod} into the kernel \eqref{eq:Meissner_mod}, we obtain the essentially same result as Eq.~\eqref{eq:tckl_kernel} which shows diamagnetic response.
While we have no staggered phase here, the additional minus sign comes from the spin-dependent velocity.
Thus we have explicitly demonstrated that the staggered nature is not the only way to stabilize OF superconductivity.

\section{N/S Junction}

In this section we consider a tunneling conductance in 
normal metal (N)/superconductor (S) junction, 
where S is a superconducting TCKL in one dimension. 
Tunneling conductance can be calculated based on the 
Blonder-Thinkham-Klapwijk theory \cite{blonder82}, and a similar method is developed also in the tight-binding system \cite{zhu83,burmistrova13}.
We choose the bulk wave functions of TCKL 
satisfying a proper boundary condition 
and calculate both the Andreev and normal reflections in this N/S junction.
For simplicity we take the half-filled case ($\mu=0$) in the following, and qualitatively same results can be obtained for $\mu\neq 0$.

\begin{figure}[t]
\begin{center}
\includegraphics[width=55mm]{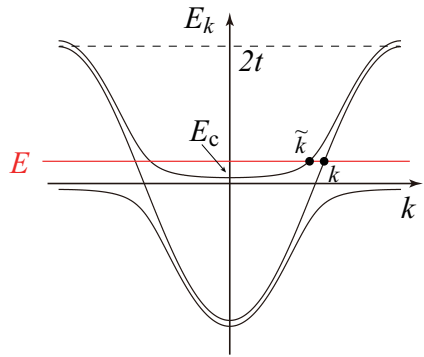}
\caption{(Color online)
Schematic illustrations for 
energy dispersion relations in one-dimensional TCKL.
}
\label{fig:dispersion}
\end{center}
\end{figure}

The mean-field Hamiltonian introduced in the previous section can be decomposed into two sets of subsystems: $(c_{k1\ua}, c^\dg_{-k-Q,2\da}, f_{k\ua})$ and $(c_{k1\da}, c^\dg_{-k-Q,2\ua}, f_{k\da})$ with $Q=\pi$.
We focus on the former set, where the eigenenergies $E_{kp}$ are given by
\begin{align}
E_{k\pm} &= \frac{\ep_k \pm \sqrt{\ep_k^2 + 8V^2}}{2}, \ \ E_{k0} = \ep_k
\end{align}
and the corresponding eigenoperators by
$\al_{kp} = u_{kp}c_{k1\ua} + v_{kp} c^\dg_{-k-Q, 2\da} + w_{kp} f_{k\ua}$ 
($p=\pm,0$) where
\begin{align}
u_{k\pm} &= -v_{k\pm} = \frac{V}{\sqrt{E_{k\mp}^2 + 2V^2}}
, \ \ w_{k\pm} = \frac{- E_{k\mp}}{\sqrt{E_{k\mp}^2 + 2V^2}}
\\
u_{k0} &= v_{k0} = 1/\sqrt 2, \ \ w_{k0} = 0
\end{align}
with $\ep_k = -2t \cos k$ and $|u_{kp}|^2 + |v_{kp}|^2 + |w_{kp}|^2 = 1$.
The dispersion relation is illustrated in Fig.~\ref{fig:dispersion}.
When we take another subsystems, 
the behaviors discussed in this section remain unchanged.
We note that the gapless part $\alpha_{k0}$ contributes to the diamagnetic Meissner kernel \cite{hoshino14-2}.
This is possible because the Fermi surface is composed of both electron and hole to form a Bogoliubov particle, as is distinct from an ordinary metal.

Now we consider the N/S junction.
The normal metal with $V=0$ and staggered OF pairing state with $V\neq 0$ in Eq.~\eqref{eq:tckl_mf} are placed at the left- ($i\leq -1$) and right-hand ($i\geq 1$) sides, respectively.
When the ($\ua, 1$) conduction electron with the energy $E>0$ is injected from the left, the wave function in N is written in the vector form as 
\begin{align}
\bm \psi_{\rm N}(i) &= 
\begin{pmatrix}
1 \\
0 \\
\end{pmatrix}
\epn^{\imu qi}
+ a
\begin{pmatrix}
0 \\
1 \\
\end{pmatrix}
\epn^{-\imu q i}
+b
\begin{pmatrix}
1 \\
0 \\
\end{pmatrix}
\epn^{-\imu q i}
.
\end{align}
The results for $E<0$ are obtained from the ones for $E>0$ by using the particle-hole symmetry.
The coefficients $a$ and $b$ correspond to Andreev and normal reflection weights, respectively.
The wave vector $q$ is determined by the condition $E = \ep_q$.
Here we have only the two components because the localized pseudofermions are decoupled in N.
A part of injected electron transmits into S,
whose wave function 
is written as
\begin{align}
\bm \psi_{\rm S}(i) &= 
c
\begin{pmatrix}
u_{k0} \\
v_{k0} \\
w_{k0}
\end{pmatrix}
\epn^{\imu k i}
+d
\begin{pmatrix}
u_{\tilde k+} \\
v_{\tilde k+} \\
w_{\tilde k+}
\end{pmatrix}
\epn^{\imu \tilde k i}
\end{align}
The wave vectors satisfy the relations $E_{k0} = E$ and $E_{\tilde k+}  =E$.
Here only the wave functions with positive group velocity appear.
We note that $\tilde k$ becomes imaginary for $E<-t+\sqrt{t^2+2V^2} \equiv E_{\rm c}$, where it exists as a quickly damping evanescent wave.

\begin{figure}[t]
\begin{center}
\includegraphics[width=80mm]{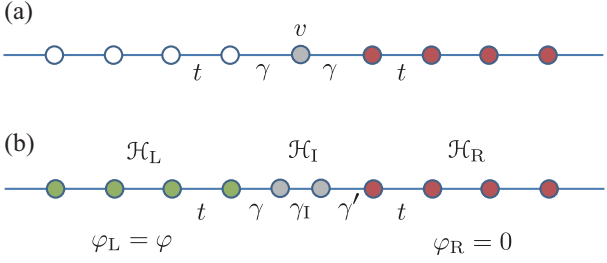}
\caption{(Color online)
Schematic illustrations for one-dimensional (a) N/S and (b) Josephson junction systems.
The on-site potential $v$ is considered at the edge of the normal metal in (a).
In (b) the Josephson current is calculated 
at the two 
% insulating 
sites located at the center of the chain.
}
\label{fig:1d_junction}
\end{center}
\end{figure}

The N part at left and the S part at right are connected at the origin by the following tunnel Hamiltonian:
\begin{align}
\mathscr{H}_{\rm I} &= 
-\gm \sum_{i=-1,0} \sum_{\sg m } (c^\dg_{i\sg m } c_{i+1,\sg m } + {\rm H.c.})
\nonumber \\
& \ \ \ 
+ v\sum_{\sg m} c^\dg_{i=0, \sg m} c_{i=0, \sg m}
. \label{eq:connect}
\end{align}
Here we consider the barrier potential $v$ at the edge of the normal metal.
The present setup of the system is schematically illustrated in Fig.~\ref{fig:1d_junction}(a).

We assume the sites for $i\leq -1$ and for $i\geq 1$ are described by the wave functions $\bm \psi_{\rm N} (i)$ and $\bm \psi_{\rm S} (i)$, respectively.
At the sites $i=-1$, $i=0$ and $i=1$ we have the relations
\begin{align}
E \bm \psi_{\rm N}(-1) &= \hat \gm \bm \psi(0) + \hat t \bm \psi_{\rm N}(-2) + 
{\hat {\mathscr{H}}_{{\rm loc},V=0}} \bm \psi_{\rm N}(-1)
\\
E \bm \psi(0) &= \hat \gm \bm \psi_{\rm N}(-1) + \hat \gm \bm \psi_{\rm S}(1) + \hat v \bm \psi (0)
\\
E \bm \psi_{\rm S}(1) &= \hat \gm \bm \psi(0) + \hat t \bm \psi_{\rm S}(2)   +
{\hat {\mathscr{H}}_{{\rm loc},V}} \bm \psi_{\rm S}(1)
\end{align}
The diagonal matrices are made from Eqs.~\eqref{eq:tckl_mf} and \eqref{eq:connect} as $\hat t={\rm diag}\,(-t,t)$, $\hat \gm={\rm diag}\,(-\gm,\gm)$ and $\hat v={\rm diag}\,(v,-v)$.
These matrices are $2\times 2$ matrices, and they operate for the upper two components of $\bm \psi_{\rm S}$, since pseudofermions have no inter-site hopping.
We also define the matrix ${\hat {\mathscr{H}}_{{\rm loc},V}}$ which originates from the local part of Eq.~\eqref{eq:tckl_mf}.
The function $\bm \psi(0)$ cannot be described in general by either $\bm \psi_{\rm N}$ or $\bm \psi_{\rm S}$ due to the presence of the potential $v$.
To determine the coefficients we need to have another relations.
This situation is similar to a usual quantum mechanics which requires smooth wave functions at the boundary.
We consider the extrapolated wave functions $\bm \psi_{\rm N}(1)$ and $\bm \psi_{\rm S}(0)$ which satisfy the relations \cite{zhu83}
\begin{align}
E \bm \psi_{\rm N}(-1) &= \hat t \bm \psi_{\rm N}(0) + \hat t \bm \psi_{\rm N}(-2)   + 
{\hat {\mathscr{H}}_{{\rm loc},V=0}} \bm \psi_{\rm N}(-1)
\\
E \bm \psi_{\rm S}(1) &= \hat t \bm \psi_{\rm S}(0) + \hat t \bm \psi_{\rm S}(2)  + 
{\hat {\mathscr{H}}_{{\rm loc},V}} \bm \psi_{\rm S}(1)
\end{align}
By solving these equations we can explicitly derive the coefficients $a$, $b$, $c$, $d$ and the wave function $\bm \psi(0)$ at the interface.

The normal reflectance $A$, Andreev reflectance $B$, and transmittances $C,D$ of quasiparticles are defined by
\begin{align}
&A = \left| \frac{\partial \ep_{- q}}{\partial (-q)}  \right|
|a|^2 \left/ \left| \frac{\partial \ep_{q}}{\partial q} \right| \right.
\label{eq:prob1}
\\
&B = \left| \frac{\partial \ep_{-q}}{\partial (-q)} \right|
|b|^2 \left/ \left| \frac{\partial \ep_{q}}{\partial q} \right| \right.
\label{eq:prob2}
\\
&C = \left| \frac{\partial E_{k0}}{\partial k} \right|
|c|^2 \left/ \left| \frac{\partial \ep_{q}}{\partial q} \right| \right.
\label{eq:prob3}
\\
&D = \left| \frac{\partial E_{\tilde k+}}{\partial \tilde k} \right|
|d|^2 \left/ \left| \frac{\partial \ep_{q}}{\partial q} \right| \right.
\label{eq:prob4}
\end{align}
which satisfy the sum rule of probability flow:
\begin{align}
A+B+C &= 1 \ \ \ {\rm for} \ \ E < E_{\rm c} \\
A+B+C+D &= 1 \ \ \ {\rm for} \ \ E > E_{\rm c} 
\end{align}
Note that the evanescent wave does not contribute to this sum rule.
From these quantities we define the conductance by
\begin{align}
\sigma(E) = 4\times \frac{e^2}{h} (1 + A - B)
, % = 1 + |a|^2 - |b|^2
\end{align}
where the factor $4$ originates from spin and orbital degrees of freedom and $h$ is the Planck constant.
The condition 
% $\sigma>1$ 
$A>B$
means the existence of an 
excess current due to  Andreev reflection, or  Cooper pair 
tunneling into the S part.
We note that in actual systems the energy is given by $E=eV_0$ with electric charge $e$ and bias voltage $V_0$.

\begin{figure}[t]
\begin{center}
\includegraphics[width=80mm]{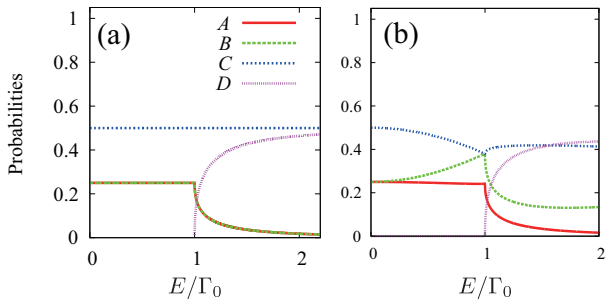}
\caption{(Color online)
Reflectances ($A,B$) and transmittances ($C,D$) as a function of energy.
The transfer integrals at the junction are chosen as (a) $\gm=t$ and (b) $\gm=0.02t$.
The barrier potential is not included in this figure.
}
\label{fig:probability}
\end{center}
\end{figure}

We first discuss the N/S junction for the $v=0$ case.
Figure \ref{fig:probability} shows the reflectances and transmittances defined by Eqs.~(\ref{eq:prob1}--\ref{eq:prob4}).
We take the hybridization strength $\Gamma_0\equiv V^2/t=10^{-4}t$, and the gap is then 
$E_{\rm c} \simeq \Gamma_0$.
In the high transmissivity limit with $\gm=t$ for $E<E_{\rm c}$ shown in Fig.~\ref{fig:probability}(a), a half of the injected electron transmits into the TCKL superconducting state ($C=1/2$).
The other half is reflected into the normal metal both as electron ($B=1/4$) and hole ($A=1/4$).
This behavior is in contrast to the ordinary 
% ESEE
$s$-wave superconductor, where the perfect Andreev reflection ($A=1$) can be observed.
For small $\gm$ case, the energy dependence is modified while the behavior at low energy remains nearly unchanged.

The presence of  normal reflection in TCKL is related to the form of the mean-field Hamiltonian given by Eq.~\eqref{eq:tckl_mf}.
Namely, the gapped structure in spectrum has the characters of both hybridization (normal) gap and superconducting (anomalous) gap.
Consequently both the normal and anomalous self energies are present as in Eqs.~\eqref{eq:normal_self} and \eqref{eq:anomalous_self}, which cause normal and Andreev reflections simultaneously.
Another characteristic behavior different from ordinary superconductors is that the transmittance into the superconducting TCKL is finite even at zero energy.
This is due to the presence of the Fermi surface as shown in Fig.~\ref{fig:dispersion}.
Hence bound states e.g. in S/N/S junction or at vortex core are unlikely formed even in the clean limit at low temperatures.

\begin{figure}[t]
\begin{center}
\includegraphics[width=85mm]{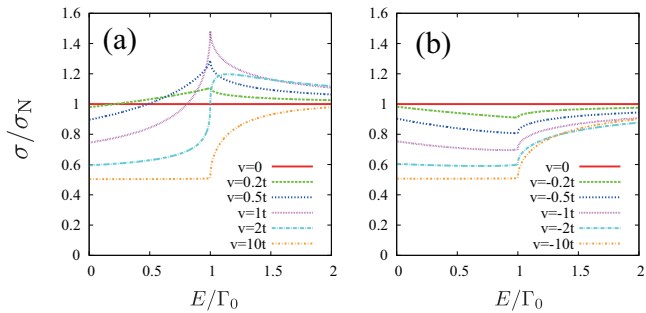}
\caption{(Color online)
Conductances for the systems with (a) repulsive ($v>0$) and (b) attractive ($v<0$) barrier potentials.
We take $\gm=t$.
}
\label{fig:conductivity2}
\end{center}
\end{figure}

Now we consider the situation with finite barrier potential at the edge of the normal metal.
The conductances are shown in Fig.~\ref{fig:conductivity2} for $\gm =t$, where we normalize them by the normal conductivity $\sg_{\rm N}=16(e^2/h)t^2\gm^2/(t^2+\gm^2)^2$.
It is characteristic that the peaked structure is observed for $v>0$ while there is no such behavior for $v<0$.
The effect of the sign of the barrier potential is remarkable near the gap edge ($E=\Gamma_0$), but it is irrelevant in the low-energy limit.

The solutions at low energies can be obtained in a simple form.
In the limit with $E\ll \Gamma_0 \ll t$, we can use the relations $q=k\simeq \pi/2 + E/2t$, $\epn^{\imu \tilde k} \simeq t E/2V^2$ and $u_{\tilde k+} \simeq E/2V$.
We then explicitly derive the reflectances for $v \ll t$ as
\begin{align}
A(E) & \sim  \frac 1 4 \left[ 1 + \left( \frac v t \right) \frac{E}{\Gamma_0} \right]
, \\
B(E) & \sim  \frac 1 4 \left[ 1 - \left( \frac v t \right) \frac{E}{\Gamma_0} \right].
\end{align}
The magnitude of the reflectance of Andreev reflection is enhanced with increasing energy $E$ for $v>0$ and is diminished for $v<0$, while the normal reflection shows the inverse behavior.
Thus the results are sensitive to the setup at the boundaries.
By contrast, the conventional spin-singlet $s$-wave superconductor does not show such a sign-sensitive behavior for barrier potential, and there is no difference between repulsive and attractive potentials.
On the other hand, for sufficiently large magnitude of potentials 
both with $v>0$ and $v<0$, the line shape of the resulting 
conductance becomes similar to 
that of the local density of states (LDOS) at the edge as will be shown in Fig.~\ref{fig:dos}(a).
This nonzero value of $\sg/\sg_{\rm N}$ at zero energy clearly characterizes 
the present superconducting state as distinct from ordinary superconductors.

The present conductance in TCKL is also different from 
that in spin-singlet $d$-wave or spin-triplet $p$-wave superconductor 
junctions \cite{Kashiwaya00}. 
In these junctions, surface Andreev bound state (SABS) produces 
a zero bias conductance peak \cite{Hu,TK95} and the magnitude of odd-frequency pairing amplitude is significant at the surface \cite{Ueda,tanaka07-2}. 
On the other hand, in the present tunneling spectroscopy of TCKL, 
the presence of the 
odd-frequency pairing does not produce a clear zero bias conductance peak.

\section{Josephson Junction}

The staggered OF pairing state of TCKL is coupled to the other 
types of superconductors in Josephson junctions.
We consider the simple spin-singlet $s$-wave superconductivity (ESEE), whose Hamiltonian is given by
\begin{align}
\mathscr{H}_{s\mathchar`-{\rm wave}}
&= 
\mathscr{H}_{\rm c} 
+ \Delta \sum_{i\sg\sg'} \epsilon_{\sg\sg'} c^\dg_{i\sg} c^\dg_{i\sg'} + {\rm H.c.}
\label{eq:s-wave}
\end{align}
in one dimension.
The conduction electron part is written as $\mathscr{H}_{\rm c}$.
As we mentioned in the introduction, there is a paramagnetic OF superconductivity which is induced only at the edge from EF superconductivity in bulk.
Here as one of such examples we take the spin-triplet $p_x$-wave superconductor (ETOE).
The Hamiltonian is explicitly written as
\begin{align}
\mathscr{H}_{p_x\mathchar`-{\rm wave}}
&= 
\mathscr{H}_{\rm c} 
+ \Delta \sum_{i\sg\sg'} \sg^x_{\sg\sg'} c^\dg_{i\sg} c^\dg_{i+1, \sg'} + {\rm H.c.}
\label{eq:px-wave}
\end{align}
When we make an edge with this Hamiltonian, the local ($s$-wave) electron pair is generated at the edge.
Since the spin structure of this pair is a triplet or even with respect to the spin exchange, the time dependence of the pair must be odd (OTEE).
This induced OF pair at the surface which is closely connected to the Andreev bound state \cite{tanaka07-2}.

The Josephson junction can be constructed by connecting TCKL with one of the above superconductors.
The Hamiltonian of this system is given by
$
\mathscr{H} = \mathscr{H}_{\rm L} + \mathscr{H}_{\rm R} + \mathscr{H}_{\rm I} 
$
each of which describes the semi-infinite left-hand system ($i\leq -1$), semi-infinite right-hand system ($i\geq 2$) and the middle junction part, 
respectively.
We take the spin-singlet $s$- or spin-triplet $p_x$-wave superconductors as $\mathscr{H}_{\rm L}$ and the TCKL as $\mathscr{H}_{\rm R}$.
The Hamiltonian for the junction part is given by
\begin{align}
\mathscr{H}_{\rm I} &=
- \sum_{\sg} \gm ( c^\dg_{i=0, \sg} c_{i=-1, \sg} + {\rm H.c.})
\nonumber \\
&\ \ 
- \sum_{ \sg} \gm_{{\rm I}} ( c^\dg_{i=0, \sg} c_{i=1, \sg} + {\rm H.c.})
+ v \sum_{i=0,1, \sg} c^\dg_{i\sg} c_{i\sg}
\nonumber \\
&\ \ 
- \sum_{ \sg m } \gm'_{\sg m} ( c^\dg_{i=1, \sg} c_{i=2, \sg m } + {\rm H.c.})
\label{eq:junction}
\end{align}
The present setup of the system in one dimension is schematically illustrated in Fig.~\ref{fig:1d_junction}(b).
The Josephson current $I$ is calculated at the center of this junction:
\begin{align}
I &= \imu e\gm_{\rm I} \sum_{\sg} \la c^\dg_{i=0, \sg} c_{i=1, \sg} - c^\dg_{i=1, \sg} c_{i=0, \sg} \ra
\label{eq:cur_def}
\end{align}
Here the Josephson  current is well defined because the gauge-symmetry breaking terms 
are not included at the junction region, 
and the equation of continuity locally holds only by quasiparticle flow.

The Josephson current can be calculated by using the semi-infinite Green function \cite{furusaki94,umerski97,yada14}.
As an alternative method one can approximate this by the Green function at the edge of the finite chain.
We take the number of sites as $N=10^5$ in the following.
The semi-infinite left- and right-hand surface Green functions $\hat g_\infty^{\rm L}(z)$ and $\hat g_\infty^{\rm R}(z)$
are explicitly derived from the Hamiltonians \eqref{eq:s-wave} and \eqref{eq:px-wave}, which can be written in a Nambu matrix form with respect to spin/orbital index.
The local Green functions at the site $i=0$ and $1$ without the connection by $\gm_{\rm I}$ are given by
\begin{align}
\hat g^{\rm L}_0 &= [z\hat 1 - \hat \gm^\dg \hat g^{\rm L}_{\infty} \hat \gm]^{-1}
, \label{eq:gL0}
\\
\hat g^{\rm R}_1 &= [z\hat 1 - \hat \gm' \hat g^{\rm R}_{\infty} (\hat \gm')^\dg]^{-1}
, \label{eq:gR1}
\end{align}
respectively.
The indices 0 and 1 mean the site index at the junction part.
The matrices $\hat \gm$, $\hat \gm_{\rm I}$ and $\hat \gm'$ are made from Eq.~\eqref{eq:junction} in a manner similar to N/S junction.
Using these quantities, the Green functions at the junction are given by
\begin{align}
\hat g_{10} &= \hat g^{\rm R}_{1} \hat \gm_{\rm I}^\dg [ (\hat g^{\rm L}_0)^{-1} - \hat \gm_{\rm I} \hat g^{\rm R}_{1} \hat \gm_{\rm I}^\dg]^{-1}
, \label{eq:g10}
\\
\hat g_{01} &= \hat g^{\rm L}_{0} \hat \gm_{\rm I} [ (\hat g^{\rm R}_1)^{-1} - \hat \gm_{\rm I}^\dg \hat g^{\rm L}_{0} \hat \gm_{\rm I}]^{-1}
.\label{eq:g01}
\end{align}
The Josephson current defined in Eq.~ \eqref{eq:cur_def} is then calculated at finite temperatures from $\hat g_{10}(\imu \ep_n)$.

\begin{figure}[t]
\begin{center}
\includegraphics[width=75mm]{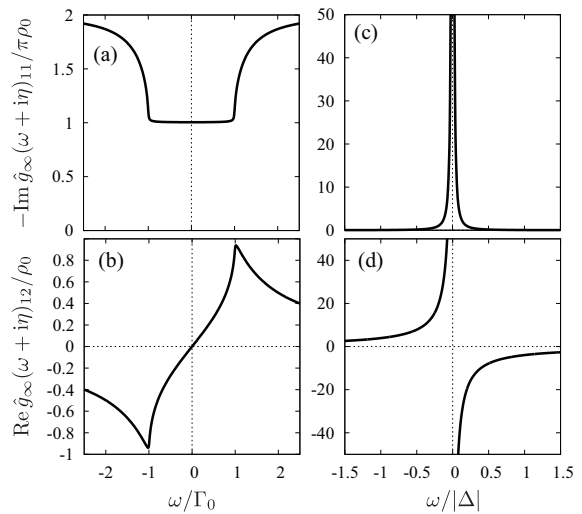}
\caption{
(a) LDOS and (b) local pair amplitude at the one-dimensional edge of the TCKL.
The same quantities are also plotted for the 
spin-triplet $p_x$-wave superconductor in (c) and (d).
The parameters are chosen as $\Gamma_0 = V^2/t=0.01t$ for (a,b) and $\Delta = 0.01t$ for (c,d).
The infinitesimal imaginary part $\eta$ is taken as $\eta = 10^{-4} t$.
}
\label{fig:dos}
\end{center}
\end{figure}

First we show the LDOS and local pairing amplitude at the edge of the semi-infinite chain calculated from $\hat g_\infty^{\rm L}$ and $\hat g_\infty^{\rm R}$.
Figures \ref{fig:dos}(a) and (b) show the LDOS  proportional 
to  $-\imag \hat g_\infty^{\rm R}(\omega + \imu \eta)_{11}$
and pair amplitude $\real \hat g_\infty^{\rm R}(\omega+\imu\eta)_{12}$, respectively, for TCKL.
The values are normalized by $\rho_0=1/2\pi t$ which is the density of states for a normal metal.
In contrast to the conventional spin-singlet $s$-wave superconductor, 
the LDOS is nonzero at the Fermi level.
This is because the half of the Bogoliubov particles in the OF pairing state 
have an energy gap and the others still have the 
Fermi surface as shown in Fig.~\ref{fig:dispersion}.
The frequency dependence of the real part of pair amplitude (or anomalous Green function) shown in Fig.~\ref{fig:dos}(b) is odd with respect to real frequency.
These behaviors are similar to the ones in bulk \cite{hoshino14-2}.
Although here we cannot see EF components, it appears as the inter-site Green functions.

On the other hand, the LDOS at the edge of the spin-triplet $p_x$-wave superconductor has the sharp peak as shown in Fig.~\ref{fig:dos}(c), which is known as a consequence of the Andreev bound state \cite{Hu,TK95,Kashiwaya00}.
This non-trivial localized edge state is formed when the sign 
of the gap function felt by quasiparticle is reversed at the reflection process.
Figure~\ref{fig:dos}(d) displays the local pair amplitude which is odd in frequency (OTEE), although in bulk only the spin-triplet $p_x$-wave EF pair (ETOE) is formed \cite{tanaka07,tanaka07-2}.

\begin{figure}[t]
\begin{center}
\includegraphics[width=75mm]{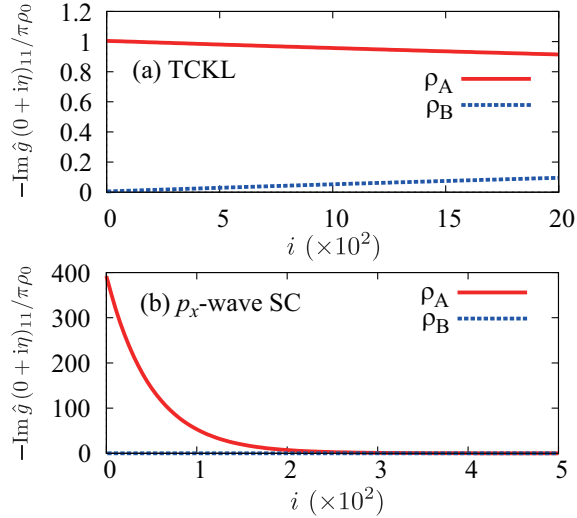}
\caption{(Color online)
Spatial dependence of LDOS at the Fermi energy.
The edge is located at $i=0$.
The parameters are same as the ones in Fig.~\ref{fig:dos}.
}
\label{fig:spat_dos}
\end{center}
\end{figure}

Let us discuss how the edge state is connected to the bulk state.
Figure \ref{fig:spat_dos} shows the spatial dependence of the LDOS at the Fermi level.
Here the LDOS at the Fermi level has the spatial dependence in the form
\begin{align}
- \imag \hat g_i (0+\imu\eta)_{11}/\pi
=
\left\{
\begin{matrix}
\rho_{\rm A}(i) & {\rm for\ even\ } i & ({\rm A\ sublattice})\\
\rho_{\rm B}(i) & {\rm for\ odd\ }  i & ({\rm B\ sublattice})
\end{matrix}
\right.
\end{align}
where $\rho_{\rm A}$ and $\rho_{\rm B}$ are smooth functions in space.
This oscillating behavior in a staggered manner originates from half-filled situation with $\mu = 0$, and the period of oscillation changes for $\mu\neq 0$ reflecting the change of Fermi wave vectors.
As shown in Fig.~\ref{fig:spat_dos}(a), the LDOS at the edge of TCKL is continuously connected to the bulk.
A slow relaxation is characteristic for the metallic state, and is consistent with the presence of Fermi surface in superconducting state of TCKL.
(In numerical simulation, the healing length, which may be physically regarded as mean free path, is given by $\sim t/\eta$ with small but finite $\eta$.)
Hence the character of this zero-energy state can be regarded as similar to the one in bulk TCKL.
For the spin-triplet $p_x$-wave superconductor shown in Fig.~\ref{fig:spat_dos}(b), on the other hand, the zero-energy state located at the edge vanishes quickly as we go into the bulk state.
This edge state has a different character from the bulk state in this case.

With these preliminaries, now we consider the Josephson junction.
In the following we consider the zero barrier potential case ($v=0$) unless explicitly stated otherwise.
The phase of the pair amplitude in the left-hand system is taken as $\varphi_{\rm L} = \varphi$, while it is set as zero in the right-hand system as illustrated in Fig.~\ref{fig:1d_junction}.
We begin with the spin-singlet $s$-wave superconductor/TCKL junction.
However, the Josephson current completely vanishes in the present simple setup.
As explained later, the absence of Josephson current is related to the fact that symmetries of the induced pairs located at the edges do not match between the left- and right-hand sides.

In order to have finite current, the simplest modification without changing bulk properties, is to change the tunnel matrix at the interface as
\begin{align}
\gm_{\sg m}' &= \left \{
\begin{matrix}
\gm_{1}' & {\rm for}\ \ (\sg m)=(\ua 1),(\da 2)
\\
\gm_{2}' & {\rm for}\ \ (\sg m)=(\ua 2),(\da 1)
\end{matrix}
\right.
\label{eq:gamma_setup_i}
\end{align}
with $\gm_{1}'\neq \gm_{2}'$.
We call this the setup (i).
Note that with this tunnel matrix both the spin and orbital symmetries are broken but their product is not broken (see also Tab.~\ref{tab:induced_pair}).
On the other hand, the more realistic setup giving finite currents is to modify the bulk nature of TCKL with keeping the tunnel matrix $\gm_{\gm\sg}' = \gm'$.
We consider the orbital field both for conduction electrons and localized pseudospin, whose Hamiltonain is given by
\begin{align}
\mathscr{H}_{\rm orb} &= -h \sum_{i\sg m} \sg^z_{mm} c^\dg_{i\sg m} c_{i\sg m}
- H \sum_{i m} \sg^z_{mm} f^\dg_{i m} f_{i m}
. \label{eq:orb_field}
\end{align}
(called the setup (ii) in the following)
This term breaks the orbital symmetry and experimentally corresponds to the uniaxial pressure effect.
When we make the junction in real materials, some stress should be applied to the edge of TCKL.
Hence the effect of Eq.~\eqref{eq:orb_field} will reasonably appear.
For simplicity we take $H=h$ in the following, but this assumption does not affect qualitative aspect of the results.

\begin{figure}[t]
\begin{center}
\includegraphics[width=85mm]{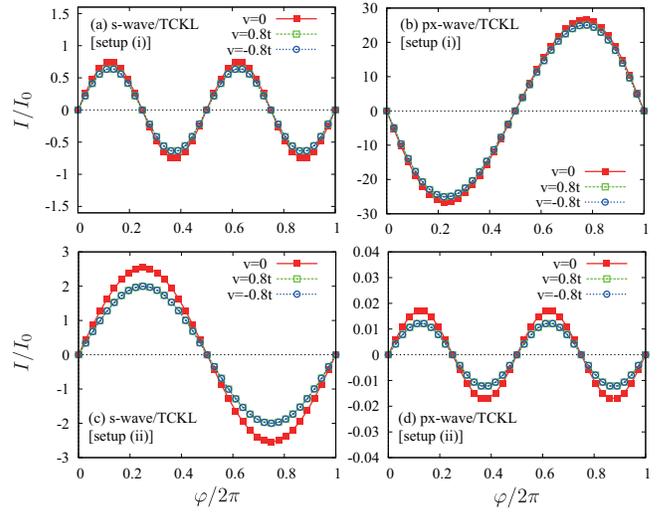}
\caption{(Color online)
Relative phase $\varphi$ dependences of the Josephson current $I$ for (a) spin-singlet $s$-wave and (b) spin-triplet $p_x$-wave superconductors connected to TCKL with the setup (i) 
[$\gm'_1=2\gm'_2=t$, $H=h=0$].
The results for the setup (ii) [$\gm'_1=\gm'_2=t$, $H=h=0.2t$] are shown in (c) and (d).
The parameters are chosen as $\Delta = 0.01t$, $\Gamma_0= V^2/t=0.01t$, $\gm=\gm_{\rm I}= t$ and $T=0.003t$.
The results for finite barrier potentials ($v=\pm0.8t$) are also shown.
}
\label{fig:1d_current}
\end{center}
\end{figure}

Figure \ref{fig:1d_current} shows the phase $\varphi$ dependence of Josephson currents, which is normalized by $I_0 = e\gm_{\rm I} (|\Delta|/\gm_{\rm I})^2$.
Let us first discuss the case with the setup (i).
As shown in Fig.~\ref{fig:1d_current}(a), the Josephson current has the functional form of $I \propto \sin 2 \varphi$ for spin-singlet $s$-wave superconductor/TCKL junction.
This indicates that the first-order coupling vanishes in this case.
The Josephson current for the spin-triplet $p_x$-wave superconductor/TCKL junction have the form $I \propto \sin \varphi$ as seen in Fig.~\ref{fig:1d_current}(b).
On the other hand, the results are changed for the setup (ii) as shown in Fig.~\ref{fig:1d_current}(c,d).
The first-order coupling survives for $s$-wave superconductor/TCKL junction, while it vanishes in the junction using $p_x$-wave superconductor.

These behaviors can be qualitatively understood by considering the two-site model (zero-dimensional system) that simulates the edges of right- and left- superconductors.
Here we focus on the case with the setup (ii), and the more detailed analysis including the setup (i) is given in Appendix.
The local spin-singlet $s$-wave pairing field term is given by Eq.~\eqref{eq:s-wave},
and the pairing field for TCKL by Eq.~\eqref{eq:tckl_mf}.
We directly connect these two sites by the matrix defined by the third line of Eq.~\eqref{eq:junction}.
The Josephson current is explicitly evaluated as
\begin{align}
I = 4 e (\gm')^2 \sin \varphi \,
T\sum_n F^{\rm L} (\imu \ep_n) \sum_{mm'} F^{\rm R}_{mm'} (\imu \ep_n)
+ O((\gm')^4	)
\label{eq:current_twosite}
\end{align}
within the lowest-order approximation.
The left- and right-anomalous Green functions are given by
\begin{align}
&
F^{\rm L}(z) = \frac{|\Delta|}{z^2 - |\Delta|^2}
= F^{\rm L}(-z)
, \nonumber \\
& 
F^{\rm R}_{mm'}(z) = \frac{\epsilon_{mm'} \Gamma_m}{(z+\sg^z_{mm} h-\Gamma_m)^2 -  \Gamma_m^2}
\end{align}
where $\Gamma_m (z) = V^2 / (z+\sg^z_{mm} H)$ is the orbital-dependent hybridization function.
For $h=H=0$ the anomalous Green function of TCKL is a purely odd function with respect to frequency, 
but the even-frequency component mixes in the presence of orbital fields.
From these expressions it is clear that the 
Josephson current becomes zero if we take zero orbital field ($h=H=0$).
With finite orbital field, on the other hand, $F^{\rm L}$ and $F^{\rm R}$ have the same parity in frequency space, and the first-order contribution to the Josephson current becomes finite.
Namely, the induced EF pair in TCKL contributes to the Josephson coupling.
For spin-triplet $p_x$-wave superconductor/TCKL junction, 
the Josephson coupling is expressed by odd-frequency spin-triplet $s$-wave 
(OTEE) and odd-frequency spin-singlet $s$-wave (OSEO) pairing. 
Then, the first-order contribution with respect to $\gm'$ vanishes.
Thus we obtain consistent results with numerical calculations for a chain discussed above.

Next we discuss the above Josephson junction from symmetry point of view.
Originally, the OSEO+ESOO pairs are present in TCKL without any field as discussed in Sec.~III.
On the other hand, for spin-singlet $s$-wave superconductor 
the ESEE pair and the induced OSOE pair are present at the edge.
In a similar manner the ETOE and induced OTEE pairs exist for spin-triplet $p_x$-wave superconductor.
Thus, no symmetries match between TCKL and the other superconductors, and the first-order coupling in Josephson junction becomes zero.
In fact, this vanishing current persists to higher orders.
To explain this behavior, we must specify the component of orbital-triplets in $s$- and $p_x$-wave superconductors.
In the present setup, since we do not include the orbital degrees of freedom explicitly, the triplet component is identified as $T_z= 1$ or $-1$ and no $T_z=0$ component.
Thus the mismatch between orbital-singlet in TCKL and orbital-triplet ($T_z=\pm 1$) in $s$- and $p_x$-wave superconductors gives exactly zero current in the present system.

With the tunnel matrix in the setup (i), the spin-orbital symmetry breaking is present and the induced pair is OTEE+ETOE according to Tab.~\ref{tab:induced_pair}.
Hence, the first-order Josephson coupling survives for $p_x$-wave superconductor/TCKL junction, but it vanishes for $s$-wave superconductor/TCKL case.
Similarly, with uniaxial pressure in the setup (ii), the ESEE+OSOE pairs are newly generated at the edge of TCKL, where orbital-triplet component include $T_z= 1$.
Hence the first-order contribution to Josephson current becomes nonzero  
for TCKL/spin-singlet $s$-wave superconductor junction.
Since we rely only on the symmetry of Cooper pairs, the above discussion should be applicable also to systems in higher dimensions.

Finally we make a comment on the effect of a barrier potential $v$ at the junction part ($i=0,1$).
The phase $\varphi$ dependence of the currents with repulsive and attractive potentials are shown in Fig.~\ref{fig:1d_current}.
The functional forms are not influenced qualitatively by the sign of $v$, since the barrier potential does not create any new species of Cooper pairs.
In addition, we do not observe the difference between $v>0$ and $v<0$.
This behavior is consistent with results in the N/S junction: the sign of the barrier potentials does not affect the behaviors in the low-energy limit as shown in Fig.~\ref{fig:conductivity2}.

\section{Summary}

We have investigated the staggered OF pairing realized in TCKL from a symmetry point of view.
Although the pair potential is purely odd function with respect to time (frequency), both OF and EF components of pair amplitude are present due to the absence of translational invariance even in the bulk.
The existing pairs in bulk are identified as primary OSEO and secondary ESOO.
We have also shown that a local gauge transformation changes the staggered state into uniform one with spin-symmetry broken state.
The mechanism for the diamagnetic Meissner effect has been explained by focusing on the symmetry of pair amplitude and structure of the Meissner kernel. 
In addition to time/spin/space/orbital structures of Cooper pairs, the finite center-of-mass momentum, which affects the sign of the velocity, plays an important role for diamagnetic response.

The N/S junction has been constructed and it is shown 
that the normal reflection is always present in addition to Andreev reflection.
This behavior is in contrast with ordinary BCS superconductors, where only the 
Andreev reflection is observed for high transmissivity limit.
The difference lies in the presence of normal self energy in the superconducting state of TCKL.
Due to a finite density of states, the transmittance into TCKL is also nonzero even at low energies.
Hence the bound state at e.g. superconducting vortex core is unlikely to be formed.
When we consider the barrier potential at the interface, the conductance shows 
the difference between attractive and repulsive potentials, 
although no such difference is observed in conventional superconductors.

We have also investigated the Josephson junction using Green function formalism. 
We connect TCKL both with spin-singlet $s$-wave and spin-triplet $p_x$-wave pairing states. 
Here a uniaxial pressure effect is considered for TCKL, which is naturally expected at the edge of real materials.
For TCKL/spin-singlet $s$-wave superconductor junction, the relative phase $\varphi$ dependence of Josephson current becomes $I\propto \sin\varphi$.
It becomes $I\propto \sin 2\varphi$ for TCKL/$p_x$-wave superconductor junction, where no first-order coupling appears.
These Josephson currents can be qualitatively described 
by a zero-dimensional system. 
The symmetry of the pairs induced at the edge is a key ingredient 
to understand the current phase relations of Josephson junctions.

\section*{Acknowledgement}
This work was financially supported by a Grant-in-Aid for JSPS Fellows (Grant No. 13J07701),
a Grant-in Aid for Scientific Research on Innovative Areas ``Topological Material Science'' (Grant No. 15H05853), and a Grant-in-Aid for Scientific Research B (Grant No. 15H03686).

\appendix

\section{Toy model analysis for Josephson junction}

We consider the simple two-site model given by $\mathscr{H} = \mathscr{H}_{\rm L}+\mathscr{H}_{\rm I}+\mathscr{H}_{\rm R}$ where
\begin{align}
&\mathscr{H}_{\rm L} = \Delta c^\dg_{{\rm L}\ua} c^\dg_{{\rm L}\da} + {\rm H.c.}
, \\
&\mathscr{H}_{\rm I} = -\sum_{\sg m} \gm'_{\sg m} c^\dg_{{\rm L}\sg} c_{{\rm R}\sg m} + {\rm H.c.}
, \\
&\mathscr{H}_{\rm R} = - h \sum_{m\sg} \sg^z_{mm} c^\dg_{{\rm R}\sg m}c_{{\rm R}\sg m}
- H \sum_{m} \sg^z_{mm} f^\dg_{{\rm R}m}f_{{\rm R}m} 
, \nonumber \\
& + V\sum_{mm'}(\delta_{mm'} f^\dg_{{\rm R}m} c_{{\rm R}\ua m'} + \epsilon_{mm'}f^\dg_{{\rm R}m} c^\dg_{{\rm R}\ua m'}) + {\rm H.c.}
\end{align}
Here $\mathscr{H}_{\rm L}$ and $\mathscr{H}_{\rm R}$ simulate the edge of the $s$-wave superconductor and TCKL, respectively.

The tunnel matrix is given by Eq.~\eqref{eq:gamma_setup_i}, and the current is simply defined by 
\begin{align}
I = \imu e \sum_{\sg m} \gm'_{\sg m} \la c^\dg_{{\rm L}\sg} c_{{\rm R}\sg m} - c^\dg_{{\rm R}\sg m} c_{{\rm L}\sg}\ra
.
\end{align}
We can solve this model analytically, which helps us understand the basic properties of Josephson junction.
We define the Green function
\begin{align}
\bm G(\tau) = 
\begin{pmatrix}
- \la T_\tau c_{{\rm L}\ua} (\tau) c^\dg_{{\rm R}\ua 1} \ra
\\
- \la T_\tau c^\dg_{{\rm L}\ua} (\tau) c^\dg_{{\rm R}\ua 1} \ra
\end{pmatrix}
.
\end{align}
The first component of this vector is relevant to current.
Its Fourier transformation $\bm G(z)$ satisfies the equation
\begin{align}
\bm G = - \hat g \hat \sg^z \gm'_1 \hat G_1
\bm c
+ \hat g \hat \sg^z \sum_m (\gm'_m)^2 \hat G_m \hat \sg^z \bm G
, \label{eq:g_calcs}
\end{align}
where $\bm c = ^{\rm t}(1,0)$ is the constant vector and
\begin{align}
\hat g &= \frac{1}{z^2- |\Delta|^2}
\begin{pmatrix}
z & |\Delta| \epn^{\imu \varphi} \\
|\Delta| \epn^{-\imu \varphi} & z
\end{pmatrix}
, \\
\hat G_m &= \frac{1}{(z+\sg^z_{mm} h-\Gamma_m)^2 -  \Gamma_m^2}
\nonumber \\
&\ \ \ \times
\begin{pmatrix}
z - \sg^z_{mm} h - \Gamma_m & \sg^z_{mm} \Gamma_m \\
 \sg^z_{mm} \Gamma_m & z + \sg^z_{mm} h - \Gamma_m
\end{pmatrix}
, \\
\Gamma_m &= V^2/(z+\sg^z_{mm}H)
.
\end{align}
The other contributions to current can also be calculated in a similar manner.
From the above equations, we can obtain Eq.~\eqref{eq:current_twosite}.

For the special case with $\gm_1'=\gm_2'=\gm'$ and $H=h=0$, namely without any symmetry breaking fields,
the Green function matrix $\sum_m \hat G_m$ becomes diagonal.
Correspondingly, the anomalous parts in the second term of the Eq.~\eqref{eq:g_calcs}, which is relevant to  higher-order Josephson couplings, are effectively dropped from the equation and the Josephson coupling terms vanish in general.
This behavior is consistent with the results discussed in Sec. V.

\end{document}